# Magnetocaloric effect and improved relative cooling power in ($La_{0.7}Sr_{0.3}MnO_3$/$SrRuO_3$) superlattices

Q. Zhang,[1] S. Thota,[1] F. Guillou,[1] P. Padhan,[2] V. Hardy,[1] A. Wahl[1] and W. Prellier,[1]

[1]Laboratoire CRISMAT, UMR 6508, CNRS ENSICAEN, 6 Boulevard du Maréchal Juin, F-14052 Caen Cedex 4, France

[2]Department of Physics, Indian Institute of Technology Madras, Chennai-600036, India

## Abstract

Magnetic properties of a series of ($La_{0.7}Sr_{0.3}MnO_3$/$SrRuO_3$) superlattices, where the $SrRuO_3$ layer thickness is varying, are examined. A room-temperature magnetocaloric effect is obtained owing to the finite size effect which reduces the $T_C$ of $La_{0.7}Sr_{0.3}MnO_3$ layers. While the working temperature ranges are enlarged, $-\Delta S_M^{max}$ values remains similar to the values in polycrystalline $La_{0.7}Sr_{0.3}MnO_3$. Consequently, the relative cooling powers are significantly improved, the microscopic mechanism of which is related to the effect of the interfaces at $La_{0.7}Sr_{0.3}MnO_3$/$SrRuO_3$ and higher nanostructural disorder. This study indicates that artificial oxide superlattices/multilayers might provide an alternative pathway in searching for efficient room-temperature magnetic refrigerators for (nano)microscale systems.

Room-temperature (RT) magnetic refrigeration [1] based on the magnetocaloric effect (MCE) has currently attracted an increasing interest because it offers an energy-efficient and environment-friendly alternative for the usual vapor-cycle refrigeration technology. In order to probe the magnetic refrigeration effectiveness, isothermal entropy alone is however, not sufficient. The relative cooling power (RCP) is indeed considered to be the most important factor for assessing the usefulness of a magnetic refrigerant material [2-3]. To date, the seek for magnetocaloric materials with high RCP was restricted mainly to polycrystalline systems or superparamagnetic nanoparticles.

In fact, the concept of magnetic refrigeration using multilayers/superlattices was introduced very recently by Mukherjee [4], who expected that the macroscopic magnetocaloric properties could be improved in nanostructural thin films with respect to corresponding polycrystalline systems. Moreover, the thin-film magnetocaloric materials may be applied in functional (nano)microscale devices for magnetic refrigeration [4-6]. Up to now, however, several attempts to study the MCE in thin films were mainly performed on single layer. Recarte *et al.* [5] investigated the Ni-Mn-Ga monolayer with first-order martensitic transformation, and suggested that the total entropy changes involving the spin entropy changes and lattice entropy changes [7], are reduced to about one third of the values exhibited in its polycrystalline system and the resultant RCP values are also decreased significantly, mainly resulting from the suppression of the martensitic transformation as well as the magneto-structural couplings in the thin film form[5]. Similar strong decrease in total entropy changes and RCP values are also found in MnAs monolayer film [6], which also involves a magneto-structural coupling around the first-order paramagnetic-ferromagnetic transition. Other investigations on monolayer films, like $La_{0.78}Ag_{0.22}MnO_3$ [8], $La_{0.67}A_{0.33}MnO_3$ (A=Ca, Sr, or Ba) [9] or $Gd_{1-x}W_x$ [10], also lead to reduced isothermal entropy changes and RCP values. Thus, it is interesting to study the MCE in the multilayers or superlattices, especially consisting of the materials with second-order transitions, *i.e.,* without the magneto-structural couplings. Furthermore, although attempts to study the MCE in metallic Gd/W[11] multilayers showed a reduced magnetic entropy changes and decreased RCP, it is well known that the synthesis of smooth and sharp (at an atomic-scale level of the order of a few Angstroms) layer interfaces in perovskite oxide magnetic multilayer structures can significantly influence the magnetic properties [12-13]. For these reasons, we have investigated the magnetocaloric properties of ($La_{0.7}Sr_{0.3}MnO_3$/$SrRuO_3$) superlattices, consisting of two perovskite systems with second-order transitions, namely $La_{0.7}Sr_{0.3}MnO_3$ (LSMO) and $SrRuO_3$ (SRO). Interestingly, when comparing with the polycrystalline $La_{0.7}Sr_{0.3}MnO_3$ compound, ($La_{0.7}Sr_{0.3}MnO_3$/$SrRuO_3$) superlattices exhibit a comparable magnetic entropy changes but a significantly improved relative cooling power. These results are discussed and solutions to overcome the intrinsic limitations of film forms are also proposed.

The ($La_{0.7}Sr_{0.3}MnO_3$/$SrRuO_3$) superlattices were grown on [001]-oriented $SrTiO_3$ (STO) substrates using a multitarget pulsed laser (KrF, $\lambda$=248nm) deposition system. The bottom layer LSMO, directly grown on the STO substrate, is fixed to be 20 unit cells while the SRO layer thickness varies with different "*n*" unit cells (*n*=1, 3 and 6). The above bilayer is repeated 15 times and finally covered with an extra LSMO layer with 20 unit cells thick LSMO layer, i.e., LSMO layer termination at both ends. The preparation method and structural details have been published elsewhere [14]. To calculate the thickness of LSMO and SRO layers, we have carried out quantitative refinement of the $\theta - 2\theta$ scan of the trilayer structures using DIFFAX program.[12,14] The high quality of the samples is also confirmed by the good agreement between the intense satellite peak positions in the $\theta - 2\theta$ x-ray diffraction patterns and the simulation profiles. Since in the literature, the $-\Delta S_M$ values are most often in unit of J/kg K, we also used this unit in the present work for the sake of comparison. Note that theoretical density values of LSMO and SRO are close to each other (6.42 and 6.39 g/cm$^3$, respectively). Moreover, the density value in film form is larger than the experimental density value in polycrystalline La$_{0.67}$Sr$_{0.33}$MnO$_3$ compound (6.23 g/cm$^3$) [15]. Thus, to avoid any overestimation of magnetocaloric properties caused by the introduction of density, we adopted the largest theoretical density value of 6.42 g/cm$^3$ for calculating the magnetic entropy changes with a unit of J/kg K. The magnetic properties were measured by applying a field along the [100] in-plane direction in a superconducting quantum interference device magnetometer (Quantum Design MPMS).

The temperature dependences of the zero-field-cooled (ZFC) and the field-cooled (FC) magnetization of the (LSMO/SRO) superlattices series recorded at 50 Oe are shown in Fig. 1. With decreasing the temperature, a sharp increase of magnetization is observed below $T_C$ = 325 K, due to the onset of ferromagnetic (FM) order in LSMO layers. This reduced $T_C$ compared to polycrystalline LSMO (365K) [16] is ascribed to the finite size effect in superlattices [4,13,14] and is found to be almost independent of n values (*i.e.* the number of SrRuO$_3$ layer). Also, the absence of thermal hysteresis around $T_C$ confirms that this transition is of second-order, in agreement with the polycrystalline LSMO [17]. At lower temperature (below 150 K), the increase of magnetization in the *n*=6 superlattice results from the formation of stoichiometric FM SRO layers [14]. In addition, below 100 K, one observes a strong decrease of the ZFC magnetization. This phenomenon is weakened for FC magnetization due to the effect of magnetic anisotropy resulting from the formation of the stoichiometric SRO layers. In the case of very thin SRO layer corresponding to superlattices with $n$ =1 and 3, both ZFC and FC magnetization significantly decrease below 100K suggesting that the stoichiometric FM SRO layers are not fully formed. The magnetization values, recorded in a field of 50 Oe below $T_C$, are much higher in our superlattices than those (≈9 emu/g) in a field of 100 Oe reported in polycrystalline LSMO[16], which is ascribed to enhanced magnetization (see below) and the in-plane direction of the easy axis [9]. When one compares to polycrystalline LSMO [16], it can also be seen that all the superlattices undergo a smoother PM-FM transition *i.e.,* with larger temperature interval between PM and FM states (named $\delta T_C$ here) due to the higher nanostructural disorder [5].

Fig. 2 displays the representative in-plane magnetic isotherms M(H) of the (LSMO/SRO) superlattices with n = 3 and 6 around $T_C$. Before measuring *M(H)* at each temperature, the sample was firstly heated up to 395 K (>$T_C$). As seen in Fig. 2, the magnetization increases rapidly at low fields and saturates at higher field values, as expected for a ferromagnetic behavior. No magnetic hysteresis is found, confirming the second-order character of the FM-PM transition. Using the Maxwell relation $(\frac{\partial S}{\partial H})_T = (\frac{\partial M}{\partial T})_H$, the magnetic-entropy change $-\Delta S$ (T) can be calculated as follow:

$$\Delta S_M(T,H) = \int_0^H (\frac{\partial S}{\partial H})_T dH = \int_0^H (\frac{\partial M}{\partial T})_H dH \qquad \text{(Eq. 1)}$$

Fig. 3 (a) and 3 (b) shows the $-\Delta S$ (T) curves of the superlattices with different number $n$ of SrRuO$_3$ unit cells for a magnetic-field change ($\Delta H$) of 50 kOe and 20 kOe, respectively. For $\Delta H$ = 50 kOe, $-\Delta S_M^{max}$ values are found to be 4.45, 4.3 and 3.07 J/kg K around 330 K, for n of 1, 3 and 6, respectively. To further evaluate the performance in terms of refrigeration efficiency, we have computed the RCP value, which depends not only on $-\Delta S^{Max}$, but also on the overall profile of $-\Delta S_M$ (T). RCP can be calculated [3] by

$$RCP = -\Delta S_M^{max} \cdot \delta T_{FWHM} . \qquad (Eq.\ 2)$$

where $\delta T_{FWHM}$ is the full width at half maximum of the $-\Delta S$ vs T curve.

From the viewpoint of applications, it is very beneficial to obtain a large magnetic entropy changes and a high RCP for $\Delta H = 20$ kOe since such a low field can be realized by using NdFeB permanent magnet. For comparison, Table I summarized for $\Delta H = 20$ kOe, the main parameters of our superlattices and polycrystalline La$_{0.7}$Sr$_{0.3}$MnO$_3$ as well as another polycrystalline La$_{0.67}$Sr$_{0.33}$MnO$_3$ with very close composition. Note that the RCP values are not sensitive to La/Sr ratio around 7/3. One of the interesting feature of Fig. 3 and Table I is that the $\Delta S_M$ (T) peaks in all investigated superlattices are significantly broadened over a wider temperature region than in corresponding polycrystalline LSMO, due to higher nanostructural disorder [5,11]. This is related to the previously noted increase in $\delta T_C$. For $\Delta H = 20$ kOe, the $\delta T_{FWHM}$ values are around 54 K when n is 1 and 3, and $\delta T_{FWHM}$ further increases significantly to 66 K for the $n=6$ superlattice. More importantly, while the working temperature ranges are enlarged, $-\Delta S_M^{max}$ values are still kept to be comparable with the values in polycrystalline La$_{0.7}$Sr$_{0.3}$MnO$_3$. It must be pointed out that thinner SRO layers ($n$=1 and 3) exhibit a higher value for $-\Delta S_M^{max}$ compared to *n*=6. The RCP values derived from Equation (2) in the superlattices are found to be significantly larger than those reported in polycrystalline LSMO.

Let us now investigate microscopic mechanisms to explore the origin of the improved RCP values in SRO-modulated (LSMO/SRO) superlattices. Assuming that only LSMO layers contribute to the magnetic entropy changes since the ordering temperature of SRO (around 150 K) is far below the investigated MCE temperature region, the maximum magnetic entropy changes $-\Delta S_{norm}^{max}$ (after normalizing to the LSMO mass only) should be almost the same. However, the normalized $-\Delta S_{norm}^{max}$ values for a relative low field change of 20 kOe are 2.51, 2.51 and 1.95 J/kg K for *n*= 1, 3 and 6 superlattices, respectively. There is no difference in the normalized $-\Delta S_M^{max}$ values for n=1 and 3, but the difference becomes obvious when n is increased to 6. Consequently, the volume ratio between LSMO and SRO is not the sole parameter that could influence the $-\Delta S_M^{max}$ of the (LSMO/SRO) superlattices around $T_C$ and other factors must be considered.

Equation (1) shows that the magnitude of $\Delta S_M$ is strongly dependent on the magnitude of $|dM/dT|$ around the magnetic transition temperature, suggesting that a MCE is generally related to two factors: the magnetization values and the temperature interval $\delta T_C$ between PM and FM states around the magnetic phase transition [18]. In previous report on the MCE in all types of thin films, the transition is spread out over a wider temperature range due to a higher nanostructural disorder, corresponding to increased $\delta T_C$ and therefore, smaller $|dM/dT|$ values. This smoothing of the transition (*i.e.* smaller $|dM/dT|$ values) leads to a strong decrease of the total entropy changes in Ni-Mn-Ga [5] and MnAs [6] monolayer involving the magneto-structural coupling and also a decrease of single magnetic entropy changes in the case of $La_{0.78}Ag_{0.22}MnO_3$ [8], $La_{0.67}A_{0.33}MnO_3$ (A=Ca, Sr, or Ba) [9], $Gd_{1-x}W_x$ [10] monolayer and Gd/W [11] multilayers without the magneto-structural coupling. In our (LSMO/SRO) superlattices, although $\delta T_C$ values are also increased, the situation is different because an additional effect comes into play. It has already been pointed out[14] that the total magnetization of (LSMO/SRO) superlattices is found to be much higher than that of polycrystalline LSMO [14]. At $T_C$, the magnetization values of superlattices with thin SRO layers ($n$=1 and 3) in a field of 20 kOe are also larger than those in the polycrystalline $La_{0.7-x}Sr_{0.3+x}MnO_3$ (x=0 and 0.03), as shown in Table I. For these $n$=1 and $n$=3 superlattices, the stoichiometric SRO layers are not formed fully. The effect of roughness and the modification of the charge states of the Mn and Ru ions at the LSMO/SRO interfaces can probably induce such an enhanced total magnetization [13,14,19,20] regardless of whether the coupled SRO layer is FM or PM state. On the other hand, the $\delta T_C$ values in superlattices with $n$=1 and 3 are larger comparing to polycrystalline $La_{0.7}Sr_{0.3}MnO_3$. As a result, the $-\Delta S_M^{max}$ values derived from Equation (1) are comparable with the largest $-\Delta S_M^{max}$ values of $La_{0.7}Sr_{0.3}MnO_3$ reported in Ref. 16. In consideration to the larger $\delta T_{FWHM}$ values, it is understood that the RCP values derived from Equation (2) for superlattices with $n$= 1 and 3 are significantly improved with respect to the polycrystalline LSMO. For superlattices with thicker SRO layer ($n$=6 in our study), the stoichiometric SRO layer starts to form and suppresses the interfacial magnetic roughness, leading to a reduced magnetization around $T_C$ relative to thinner SRO layers, as can be seen from the Fig. 2 and also in Table I. Thus, compared with $n$=1 and 3 superlattices, the decreased magnetization and a slightly increased $\delta T_C$ around $T_C$ for $n=6$ superlattice, lead to a smaller $-\Delta S_M^{max}$ value, but the RCP is mostly compensated by a larger $\delta T_{FWHM}$, also resulting in an improved RCP relative to polycrystalline LSMO.

In order to exploit the superlattice potential in terms of application, different strategies could be proposed to limit the drawback of the substrate and to make use of it. For instance, apart from the reduction of its thickness, the substrate could be used to integrate micro-magnetocaloric processes within the micro-electronic circuitry. Indeed, the substrate might be patterned to make circulating in it micro-channels required for heat transfers in any micro-refrigerating devices. Superior to the (doped) monolayer film, the stack of the multilayers/superlattices possesses a larger mass (volume) of active material which could be more suitable for possible application in (nano)micro-scale refrigeration systems since miniaturization permits the magnetic cooling powers only for small objects[4]. The idea would be to increase the mass of magnetically active material without recovering bulk properties since the interfacial properties, as the origin of the improved cooling power, are intrinsic features of such systems. In addition, the use of a small amount of active material have the advantage of showing a smaller relaxation time that the heat exchange process could take to reach the stationary state in the case of macro-systems. The reduction of micro-scale allows the refrigeration cycle frequency by about a factor 10 [21].

In conclusion, we reported different magnetic and magnetocaloric effect in (La$_{0.7}$Sr$_{0.3}$MnO$_3$/SrRuO$_3$) superlattices with respect to polycrystalline La$_{0.7}$Sr$_{0.3}$MnO$_3$ compound. The transition from PM to FM states in all superlattices occurs in a wider temperature region, resulting in an enlarged working temperature region. However, the modification of the charge states of the Mn and Ru ions at the LSMO/SRO interfaces enhanced the magnetization around $T_C$, which counterbalances the negative effect of the transition broadening and leads to a comparable $-\Delta S_M^{max}$ values. The RCP values are found to be improved significantly due to the comparable $-\Delta S_M^{max}$ values and increased $\delta T_{FWHM}$ values. With the increase of $n$ from 1 to 3, the reversible $-\Delta S_M^{max}$, the large $\delta T_{FWHM}$ and high RCP value for $\Delta H = 20$ kOe are changed slightly and found to be around 2.3 J/kg, 53 K and 120 J/kg, respectively. When $n$ increased to 6, the reversible $-\Delta S_M^{max}$ decreased to be 1.52 J/kg K, whereas $\delta T_{FWHM}$ is increased to be 66 K, also resulting in a large RCP of 100 J/kg. The study on (La$_{0.7}$Sr$_{0.3}$MnO$_3$/SrRuO$_3$) superlattices, might be a stimulus to search for suitable materials with significantly improved relative cooling power in perovskite multilayers or superlattices by adjusting the interfaces and nanostructure, for the RT magnetic refrigerant in (nano)microsystems.

**Acknowledgement**

This work has been supported by the European project "SOPRANO" under Marie Curie actions (Grant. No. PITN-GA- 2008-214040), the CEFIPRA/IFPCAR (3908-1), the C'Nano Nord Ouest (GDR 2975) and STAR (21465YL), ENERMAT, the Indo-French Associated International Laboratory (LAFICS), and CNRS "Energy" program "PR Réfrigération Magnétique". We also thank J. Lecourt, L. Gouleuf.

Table I . Comparison of the main parameters of (LSMO/SRO) superlattices with those of the polycrystalline $La_{0.7}Sr_{0.3}MnO_3$ and $La_{0.67}Sr_{0.33}MnO_3$ for $\Delta H = 20$ kOe. N1, N3 and N6 denote the (LSMO/SRO) superlattices with n=1, 3 and 6, respectively.

| Material | $-\Delta S_M^{max}$ (J/kg K) | $\delta T_{FWHM}$ (K) | RCP (J/kg) | Volume ratio of LSMO (%) | Normalized $-\Delta S_{norm}^{max}$ (J/kgK) | $T_C$ (K) | M($T_C$,H=20kOe) (emu/g) | Evaluated from Ref. |
|---|---|---|---|---|---|---|---|---|
| N1 | 2.35 | 53 | 125 | 93.5 % | 2.51 | 325 | 41.6 | this work |
| N3 | 2.2 | 54 | 119 | 87.6 % | 2.51 | 325 | 42.7 | this work |
| N6 | 1.52 | 66 | 100 | 77.9 % | 1.95 | 325 | 33.8 | this work |
| $La_{0.7}Sr_{0.3}MnO_3$ | 2.66 | 26 | 69 | | | 365 | 34 | 16 |
| $La_{0.7}Sr_{0.3}MnO_3$ | 1.78 | 43 | 77 | | | 374 | | 22 |
| $La_{0.7}Sr_{0.3}MnO_3$ | 1.27 | 22.8 | 29 | | | 370 | | 23 |
| $La_{0.67}Sr_{0.33}MnO_3$ | 2.02 | 40 | 80 | | | 370 | 33 | 24 |

Figure Captions.

Fig. 1. (Color online) Temperature dependences of the ZFC (open symbols) and FC (solid symbols) in-plane magnetization of (LSMO/SRO) superlattices with n=1 (squares), 3 (triangles) and 6 (circles) in a field of 50 Oe.

Fig. 2. (Color online) In-plane magnetic isotherms of (LSMO/SRO) superlattices with (a) n=3 and (b) n=6, measured with increasing field (open squares) and decreasing field (solid triangles) processes around $T_C$.

Fig. 3. Temperature dependences of - $\Delta S_M (T)$ in (LSMO/SRO) superlattices with n= 1, 3 and 6 for $\Delta H$ of (a) 50 kOe and (b) 20 kOe, respectively.

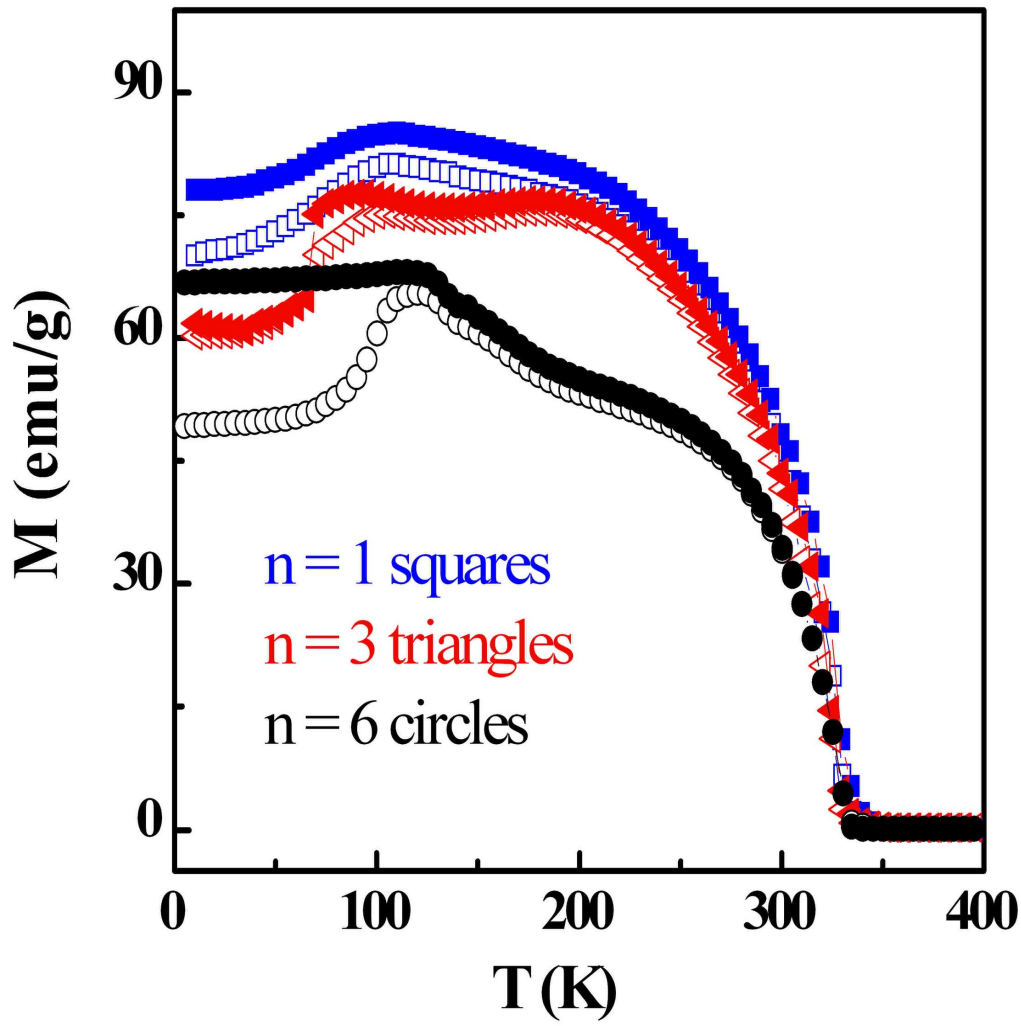

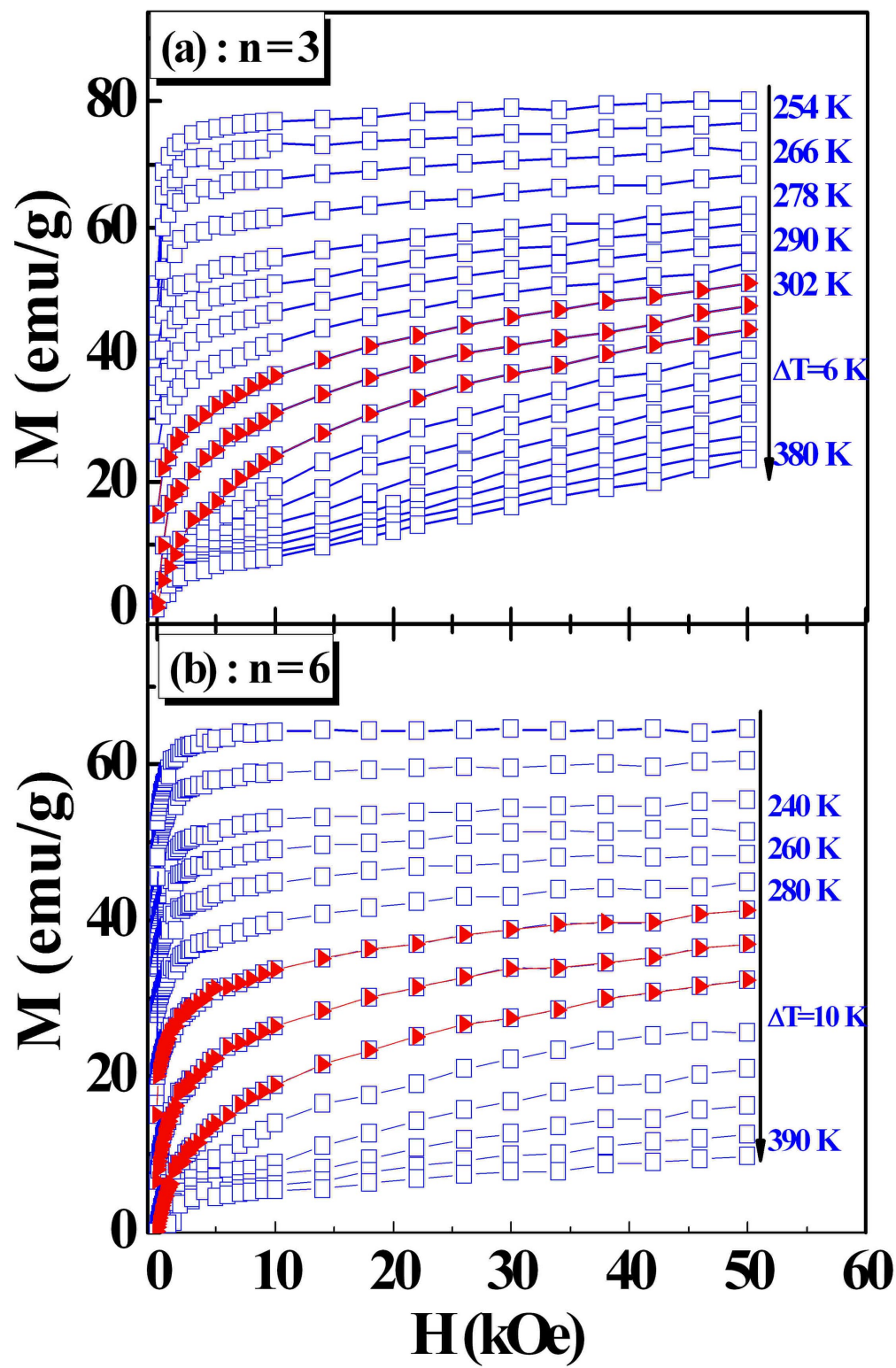

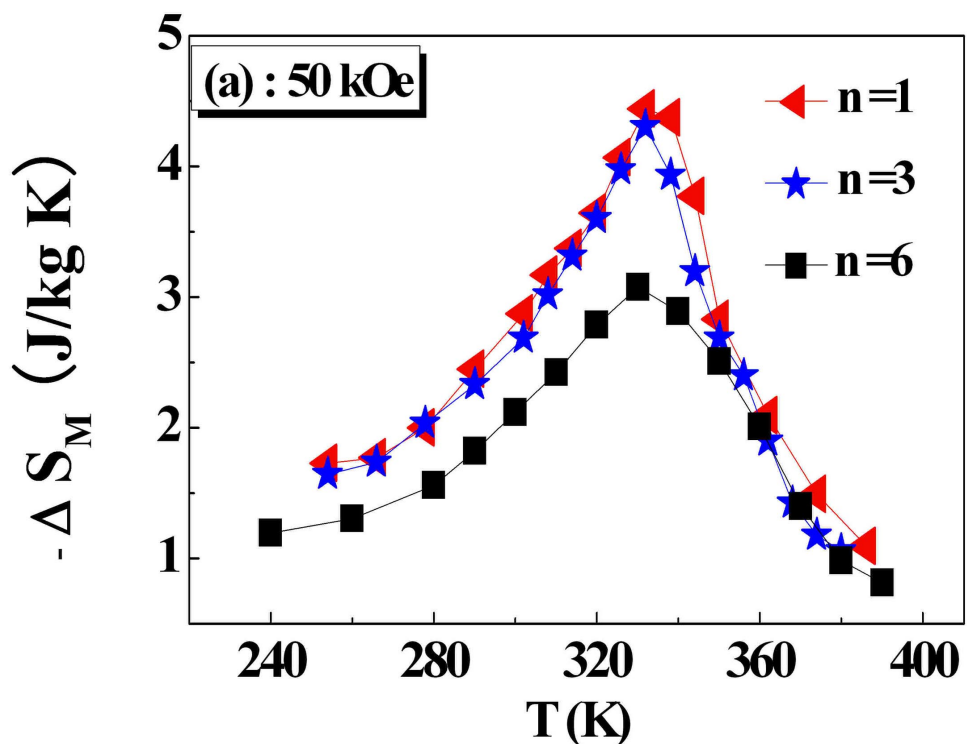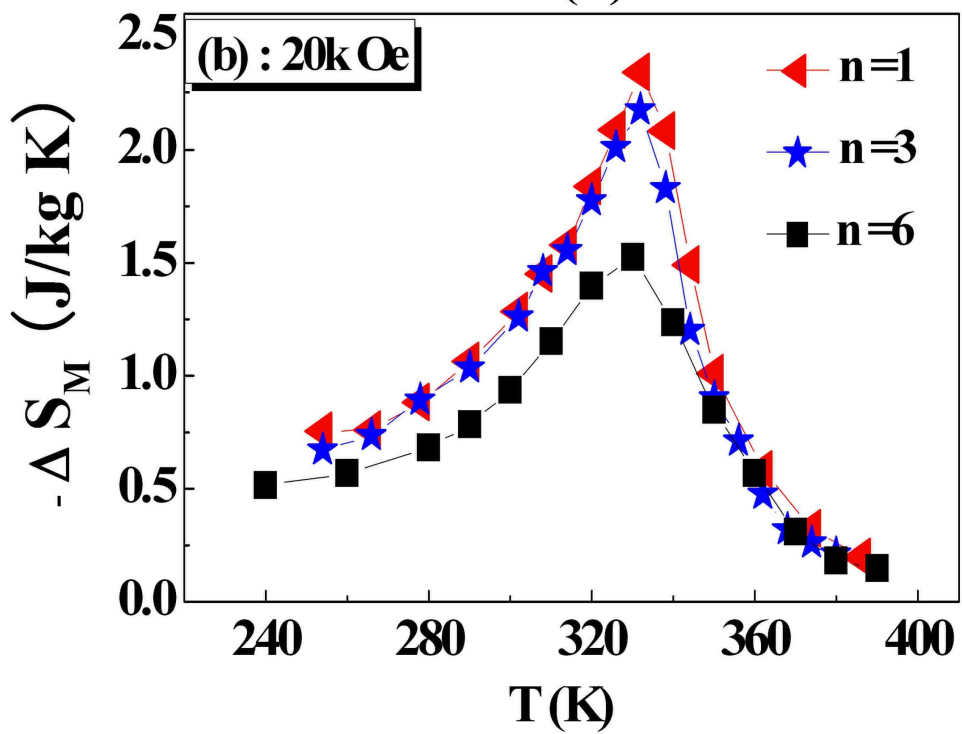